# The Sky Is Not the Limit: LTE for Unmanned Aerial Vehicles

Xingqin Lin, Vijaya Yajnanarayana, Siva D. Muruganathan, Shiwei Gao, Henrik Asplund,

Helka-Liina Määttänen, Mattias Bergström, Sebastian Euler, and Y.-P. Eric Wang

Ericsson

Contact: xingqin.lin@ericsson.com

*Abstract*— **Many use cases of unmanned aerial vehicles (UAVs) require beyond visual line-of-sight (LOS) communications. Mobile networks offer wide area, high speed, and secure wireless connectivity, which can enhance control and safety of UAV operations and enable beyond visual LOS use cases. In this article, we share some of our experience in Long-Term Evolution (LTE) connectivity for low altitude small UAVs. We first identify the typical airborne connectivity requirements and characteristics, highlight the different propagation conditions for UAVs and mobiles on the ground with measurement and ray tracing results, and present simulation results to shed light on the feasibility of providing LTE connectivity for UAVs. We also present several ideas on potential enhancements for improving LTE connectivity performance and identify fruitful avenues for future research.**

## I. INTRODUCTION

Much of the past research and development of mobile broadband communication has been primarily devoted to terrestrial communication. Providing tetherless broadband connectivity for unmanned aerial vehicles (UAVs) is an emerging field. UAVs come in various sizes, weights, and fly at different speeds and altitudes. In this article, we focus on low altitude small UAVs. The Federal Aviation Administration (FAA) guidelines may be used as a working definition for UAVs in this category: UAVs with weight no more than 55 pounds, maximum speed of 100 miles per hour, and maximum altitude of 400 feet above ground level (AGL) or within 400 feet of a structure if higher than 400 feet AGL [1]. The use cases of commercial UAVs (a.k.a. drones) are growing rapidly, including delivery, communications and media, inspection of critical infrastructure, surveillance, search-and-rescue operations, agriculture, wildlife conservation, among others [2]. For example, Amazon started Prime Air drone delivery trials in the U.K. in December 2016 and in the U.S. in March 2017 [3].

Many commercial small UAVs today are equipped with Wi-Fi connectivity so that they are remotely accessible. Wi-Fi connectivity, however, may not be sufficient for beyond visual line-of-sight (LOS) communications needs, particularly those require wide-area connectivity. To enable beyond visual LOS unmanned aircraft systems, a collaborative initiative driven by FAA and National Aeronautics and Space Administration (NASA) has been initiated in the U.S. since January 2017 [4]. One of the focus areas is communications and navigation, targeting exploring operator solutions to ensure safe control of UAVs beyond visual LOS.

Mobile networks offer wide area, high speed, and secure wireless connectivity, which can significantly enhance control and safety of small UAV operations and enable beyond visual LOS use cases. The rapid and vast growth in the small UAV industry will bring new promising business opportunities to mobile operators. Not surprisingly, the recent two years have seen a surge of activities in utilizing established Long-Term Evolution (LTE) networks for UAVs. There have been increasing field trials involving using terrestrial LTE networks to provide connectivity to UAVs [5]. More recently, telecommunications operator KDDI and Terra Drone announced the completion of a 4G LTE control system that allows operators to control small UAVs via LTE network [6].

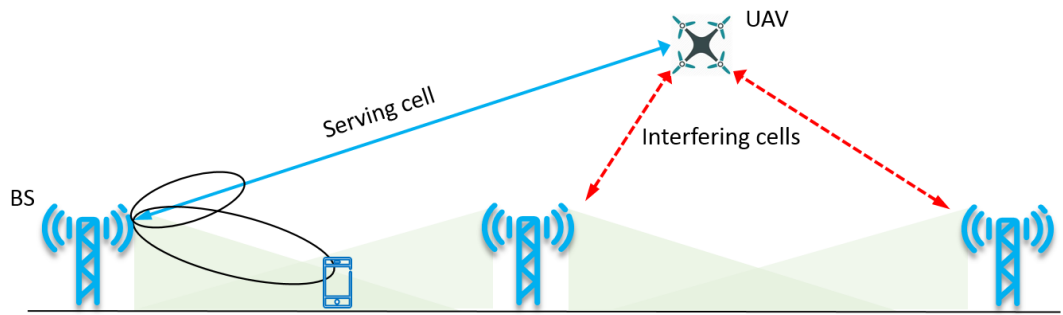

Figure 1: Illustration of wide-area wireless connectivity for low altitude small UAVs with terrestrial cellular networks.

To better understand the potential of LTE for small UAVs, the 3rd Generation Partnership Project (3GPP) has started a study item on enhanced LTE support for aerial vehicles since March 2017 [7]. In this article, we share some of our experience in this emerging field. We first identify the typical connectivity requirements and characteristics for low altitude UAVs, highlight the different propagation conditions for UAVs flying in the sky and user equipment (UE) on the ground, and present simulation results to shed light on the feasibility of providing LTE connectivity for small UAVs. We also present several ideas on potential enhancements for improving skyward LTE connectivity performance and conclude by pointing out some fruitful avenues for future research.

## II. CONNECTIVITY REQUIREMENTS AND CHARACTERISTICS

While the use cases of UAVs are many, the wireless connectivity serves two main purposes.

- *Command and control:* The ability of remote command and control can significantly enhance the safety and operation of UAVs. To ensure proper operational control of the UAVs, 3GPP requirements on command and control are data rates up to 100 kbps and packet error rate lower than 0.1% within 50 ms latency bound [9].
- *Data communication:* Use cases such as flying cameras and remote surveillance require UAVs to send back real-live telemetry data, pictures, or videos. The main connectivity requirement of such data communication is data rate, which may be up to 50 Mbps [9].

There are two main aspects that make using LTE networks for serving UAVs challenging and interesting. First is the coverage. Mobile LTE networks are optimized for terrestrial broadband communication, and thus base station (BS) antennas are down-tilted to reduce the interference power level to other cells. As illustrated in Figure 1, with downtilted BS antennas, small UAVs may be served by the sidelobes of BS antennas. The propagation conditions, however, are more favorable for UAVs flying in the sky than terrestrial propagation. These facts naturally raise the question about whether the more benign propagations can make up for antenna gain reductions.

The second aspect is the interference. With more favorable propagation conditions in the sky, small airborne UAVs may generate more uplink interference to the neighbor cells while experiencing more downlink interference from the neighbor cells. With the presence of a large number of UAV connections, the increased uplink interference, if not properly controlled and managed, may cause performance degradation to the UEs on the ground. Understanding the impact of interference is one of the key objectives of the 3GPP study item on enhanced LTE support for aerial vehicles [7].

The characteristics of coverage and interference associated with LTE radio links for UAVs are by and large different from terrestrial LTE connectivity. Besides exploring the feasibility, it is important to explore potential enhancements to provide more effective and efficient LTE connectivity for small UAVs without negatively impacting the performance of ground UEs.

## III. AERIAL CHANNEL CHARACTERISTICS

The distinct connectivity phenomena in providing LTE connectivity for UAVs is rooted in the different wireless channels. A complete characterization of the wireless channels between ground BSs and airborne UAVs is outside the scope of this article. In this section, we highlight two key aspects of the aerial wireless channels: LOS vs. non-LOS (NLOS), and large-scale pathloss.

### A. LOS versus NLOS propagation

In this article, LOS propagation is defined as a condition where the direct ray between two points is clear of obstacles. NLOS propagation is a condition where the direct ray between two points is obstructed by obstacles. At a given instant a radio link either has LOS or NLOS. The characteristics of LOS and NLOS wireless channels are dramatically different.

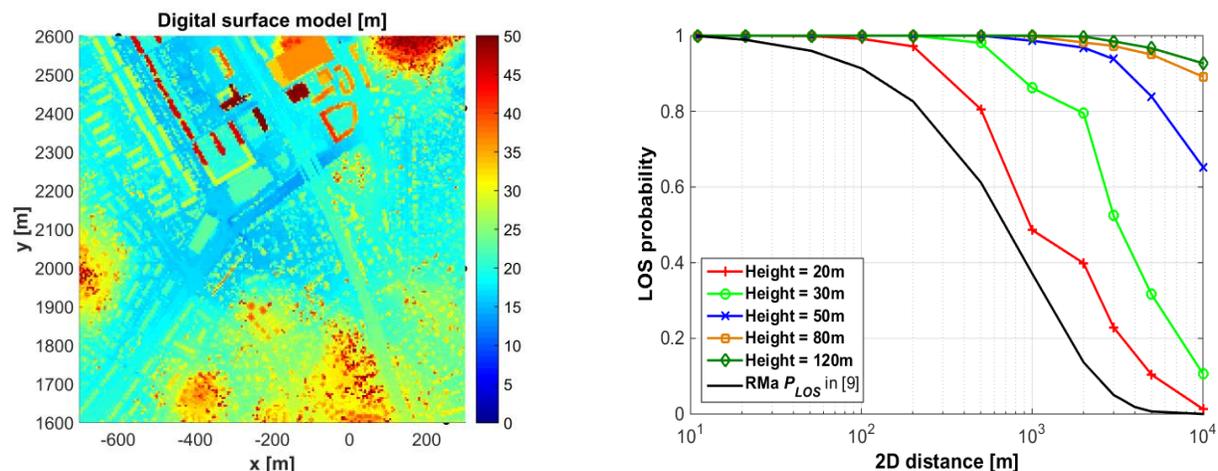

Figure 2: LOS probabilities in rural scenario: The left subfigure shows a map of a rural area near Stockholm, Sweden, used for deriving LOS probabilities. The x-y coordinates are in units of meters and the density map shows the heights in unit of meters. The right subfigure shows the LOS probabilities obtained from the rural area map data.

In system level simulations, the propagation condition of a radio link, LOS or NLOS, is usually determined according to a LOS probability function, which may depend on the distance and heights of transceivers (see e.g. Section 7.4.2 in the technical report [8] of 3GPP's study on channel models). With the determined LOS or NLOS propagation condition, the corresponding pathloss and small scale fading are generated accordingly. In our study, the effect of the Fresnel zone is not included in LOS modeling but is taken into account in pathloss modeling.

To simulate the system wide performance of LTE for UAVs, it is important to accurately model the LOS and NLOS propagation conditions. Unfortunately, the LOS probability models in [8] are not applicable to the altitudes of interest for small UAVs. For example, the maximum applicable UE height is 23 m in the models in [8].

One approach to deriving LOS probability models is ray tracing. Figure 2 shows an example of our work in this direction. The left subfigure of Figure 2 shows a high-resolution digital terrain map of a rural area near Stockholm, Sweden. The map is based on aerial 3D laser-scanned data and represents the terrain and building heights with a resolution of 5 m in the horizontal plane and 0.15 m in the vertical plane. In the rural area map, the BSs are randomly dropped at 100 different locations not occupied by the buildings. The BS height is 35 m above the terrain. For each random BS drop the antenna orientation is chosen uniformly between 0 and 360 degrees. Then the UEs are randomly dropped, and the LOS or NLOS propagation conditions are determined by examining the eventual presence of blocking buildings or terrain features for different UE heights. The right subfigure of Figure 2 shows the LOS probabilities versus 2D distance obtained from the rural area map data. Here, 2D distance of two points $(x_1, y_1, z_1)$ and $(x_2, y_2, z_2)$ in three-dimensional space is defined as the distance between the two points $(x_1, y_1)$ and $(x_2, y_2)$ in two-dimensional space. In Figure 2, the LOS probability formula from [8] is also shown for comparison. From the results, it is evident that even for heights above the BS height of 35m, the LOS probability can be less than 1. For instance, at a 2D distance of 10 km and a UE height of 50 m, the LOS probability is only 65% in the map evaluated, due to terrain height variations, etc.

LOS probability models are currently under development in 3GPP and will be captured in the technical report [9].

### B. Large-scale pathloss

Large-scale pathloss is one of (if not) the most prominent factors in estimating the received signal power for wireless systems. Statistical pathloss models are of importance for system analysis and simulation. Extensive empirical measurements have been carried out in the past few decades to develop statistical pathloss models in typical terrestrial wireless environments. Some of these statistical pathloss models adopted by 3GPP may be found in [8].

As in the case of LOS probability models, the statistical pathloss models in [8] are also not applicable to the altitudes of interest for small UAVs. For example, the rural macrocell pathloss model defined in [8] is only applicable for UE height below 10 m and 2D distance up to 10 km. For airborne UAVs at the altitudes above a certain altitude, it may be conjectured that the propagation condition is close to free space since the LOS probability is close to one and the propagation environment in the sky is clear. For the intermediate altitudes, large-scale pathloss characterization is more complicated. Figure 3 compares the 3GPP rural macrocell LOS and NLOS pathloss models for ground UEs in [8] to our measurement data collected in a helicopter measurement campaign [10]. The benchmark free-space pathloss is also shown for comparison. Since we did not distinguish LOS and NLOS data in the measurement, we expect that for reasonable LOS and NLOS pathloss models, most of the measurement data should fall in between the two models. This property, however, is not satisfied by the 3GPP rural macrocell pathloss models in [8], as shown in Figure 3.

While LOS pathloss is typically close to or lower bounded by the free-space pathloss at shorter distances, at longer ranges the curvature of the earth causes an increased loss due to diffraction. The 3GPP rural macrocell LOS model in [8] was developed for ground UEs with heights below 10 m. For airborne UEs above 10 m from the ground, using the existing 3GPP model would lead to over-estimated pathloss, particularly for UEs at higher altitudes and at large 2D distances.

Statistical pathloss models are also under development in 3GPP and will be captured in the technical report [9].

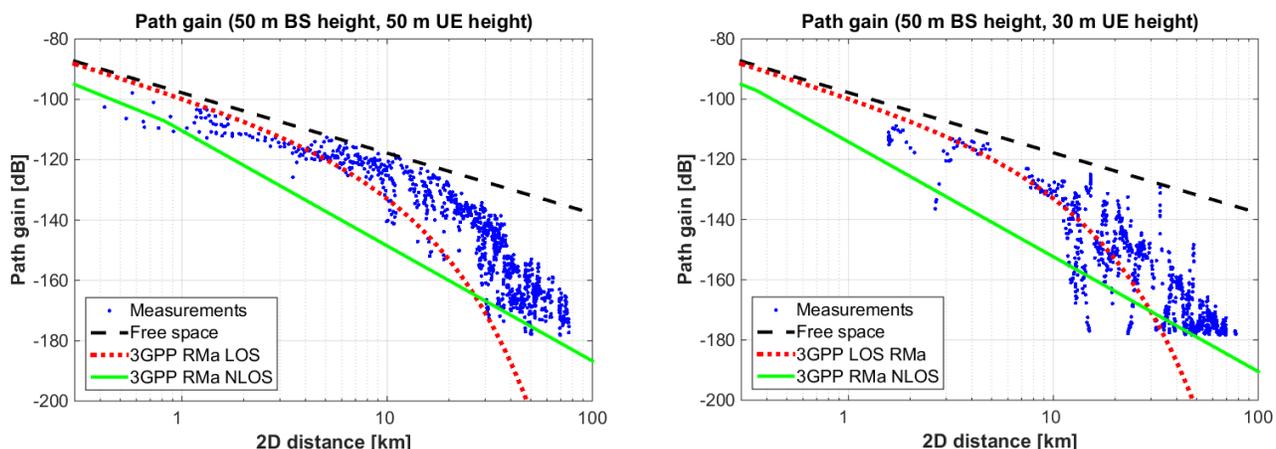

Figure 3: Comparison of 3GPP rural macro (RMa) pathloss models and measurement data: The BS height is about 50 m and the height of the surrounding clutters (trees, buildings, etc.) is about 25 m. The UE height is about 50 m above the ground in the left subfigure and 30 m in the right subfigure. The carrier frequency is 1.8 GHz.

## IV. TECHNICAL FEASIBILITY AND CHALLEGNES

In this section, we present initial simulation results to shed light on the feasibility of providing LTE connectivity for small UAVs. We consider a rural scenario, where sites are placed on a hexagonal grid with 37 sites and 3 cells per site. The LTE system bandwidth is 10 MHz at 700 MHz carrier frequency. Each BS has two cross polarized antennas with 6 degrees of downtilt at the height of 35 m.

The evaluation assumption on BS antenna pattern is important, since the airborne UAVs may be served by the sidelobes of downtilted antennas. To model the sidelobes of the antenna pattern of 2 cross polarized antennas at the BS, we synthesize an antenna pattern using an antenna array with (M, N, P) = (8, 1, 2) and $0.8\lambda$ vertical antenna element spacing, where M denotes the number of rows in the array, N denotes the number of columns in the array, P denotes polarization, and $\lambda$ denotes the wavelength. The synthesized BS antenna pattern is shown in Figure 4, which illustrates the sidelobes used for the evaluation. For aerial channel models, we reuse the 3GPP channel models in [8] for UAVs at altitudes below BS antenna height and adopt free-space propagation for UAVs at altitudes above BS antenna height.

Figure 5 shows the downlink coupling gain (antenna gain + path gain) and downlink signal-to-interference-plus-noise ratio (SINR) distributions at three different altitudes: 1.5 m (ground level), 40 m (5 m above the BS antenna height), and 120 m (close to the FAA altitude limit of 400 feet for small UAVs [1]). For a given altitude, all the UEs are placed at the same altitude. Due to downtilted BS antennas, UEs at 40 m and 120 m are served by the sidelobes of BS antennas, which have reduced antenna gain compared to the mainlobes of BS antennas serving UEs at 1.5 m. However, UEs at 40 m and 120 m have free-space propagation conditions, while radio signals attenuate more quickly with distance on the ground. From the distributions of downlink coupling gain (that combines antenna gain and channel gain) in the left subfigure of Figure 5, we can see that that the free-space propagations can make up for the BS antenna sidelobe gain reductions. In particular, the fifth percentile downlink coupling gains at the altitude of both 40 m and 120 m are higher than the fifth percentile downlink path gain at the ground level of 1.5 m.

From the SINR distributions in the right subfigure of Figure 5, we can see that the SINRs at the altitude of both 40 m and 120 m are statistically lower than the SINRs at the ground level of 1.5 m. Specifically, at the operating point of 20% resource utilization, the median SINRs at the altitude of 40 m and 120 m are 10.9 dB and 11.3 dB lower than the median SINR at the ground level, respectively. These results show that the free-space propagations also lead to stronger interfering signals from non-serving cells to the airborne UAVs.

In a Release-12 LTE network, downlink coverage normally requires a minimum SINR of -6 dB. Thus, as indicated by Figure 5, aerial UEs may be out of coverage due to interference. Coverage enhancement features were introduced in LTE Release 13, which allows a UE with SINR as low as -10 dB to be in coverage. These coverage enhancement features can be applied to aerial UEs, particularly for broadcast and common control channels.

Figure 6 shows the uplink resource utilization and uplink throughput versus total traffic per cell at different altitudes. The ratio of the number of aerial UEs to the total number of UEs is 10% in the simulation. To get baseline results, we treat the uplink traffic from aerial UEs and terrestrial UEs equally in the scheduling and apply the same uplink power control parameters regardless of the types of the UEs. The scenario with 1.5 m altitude is the benchmark, since aerial UEs are at the ground level in this scenario and thus all the UEs can be considered as "terrestrial" UEs. We now compare the performance at the ground level to the scenarios with aerials UEs at the altitude of 40 m or 120 m. It can be seen from Figure 6 that for serving the same traffic per cell, the resource utilizations are higher and the throughputs are lower when aerial UEs are flying in the sky. The aerial UEs, when at the height of 40 m and 120 m, are above the typical heights of the terrain obstacles such as buildings and trees. This results in good propagation conditions for aerial UEs to neighbor BSs, thus creating more uplink interference to neighbor BSs. How to manage and mitigate the increased interference is discussed in Section V-A.

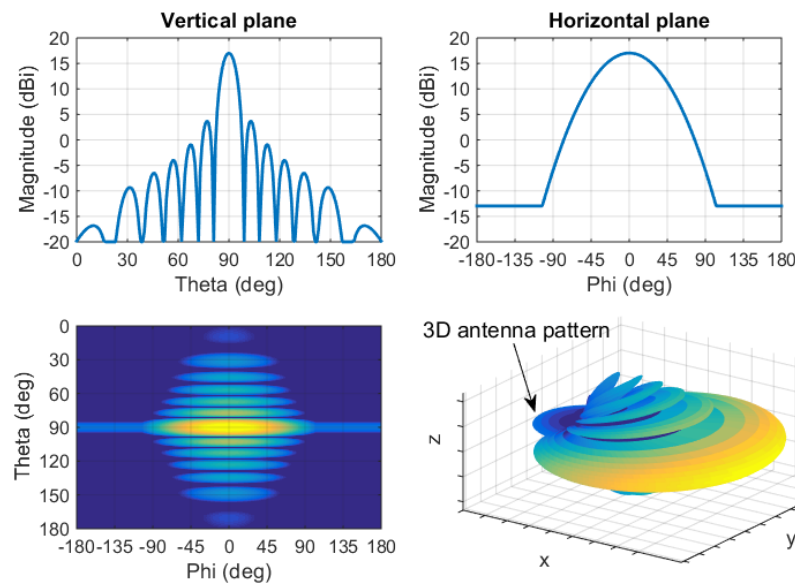

Figure 4: Synthesized BS antenna pattern for system level simulation of airborne LTE connectivity: theta denotes zenith angle, and phi denotes azimuth angle.

## V. POTENTIAL ENHANCEMENTS FOR EFFICIENT UAV CONNECTIVITY

In this section, we discuss performance enhancing solutions to optimize LTE connectivity to provide improved performance for small UAVs while protecting the performance of ground mobile devices.

### A. Interference mitigation

Inter-cell interference is not a new issue in mobile networks. A rich set of tools in terms of both standards and implementation have been studied and developed for LTE to deal with interference. In this section, we briefly discuss some prominent interference mitigation techniques. The 3GPP study item on enhanced LTE support for aerial vehicles [7] has identified interference mitigation as a key objective. The technical report [9] will capture the detailed description of interference mitigation techniques. We refer interested readers to [9] for more details.

One prominent interference mitigation tool is coordinated multipoint (CoMP) transmission and reception (and its variants) [11]. The new challenge here is that aerial UEs receive interfering signals from more ground BSs in the downlink and their uplink signals are visible to more cells due to more LOS propagation conditions. Thus, the methods need to scale for a large set of cells without much complications due to the requirements on additional pilots, synchronization, scheduling, etc. The optimal grouping strategy for cooperation among BSs, the tradeoff between overhead, coordination complexity and interference mitigation gain are still open problems.

Interference can also be handled by receiver techniques such as interference rejection combining and network-assisted interference cancellation and suppression [12]. Comparing the sizes of small UAVs to the sizes of smart phones, it is more feasible to equip small UAVs with more antennas, which can be used to cancel or suppress the interfering signals from more ground BSs. With multiple antennas, beamforming that enables directional signal transmission or reception to achieve spatial selectivity is also an effective interference mitigation technique.

A simpler interference mitigation solution would be to partition radio resources so that aerial traffic and terrestrial traffic are served with orthogonal radio resources. The static radio resource partition may not be efficient since the reserved radio resources for aerial traffic may be underutilized. If UAV operators can provide supplemental data such as flight routes and UAV positions to the network operators, such data can be utilized for more dynamic and thus more efficient radio resource management.

Uplink power control is yet another powerful interference mitigation technique. In the simulation results presented in the previous section, the same uplink power control parameters are applied regardless of the types of the UEs. An optimized setting of uplink power control parameters may be applied to limit the excessive uplink interference generated by UAVs. Optimized uplink power control can reduce interference, increase spectral efficiency, and benefit aerial UEs as well as terrestrial UEs.

One other option may be to use dedicated cells for the UAVs where the antenna patterns are pointed towards the sky instead of down-tilted. These dedicated cells will be particularly helpful in UAV hotspots where frequent and dense UAV takeoffs and landings occur.

### B. Mobility enhancement

At the ground level, the strongest site is usually the closest one. The terrestrial UEs served by a BS with a sector antenna are clustered in a contiguous area close to the serving BS. For aerial UEs, a farther BS rather than the closest BS may possibly be chosen as the serving BS. This can occur because the mainlobes of BS antennas are tilted downwards to optimize terrestrial coverage. For aerial UEs in the sky, they may be served by the sidelobes of BS antennas. Depending on the position of an aerial UE, the aerial UE may be located in the antenna null of a nearby BS and thus its received power from a farther BS may be stronger. As a result, the association pattern becomes fragmented in the sense that aerial UEs served by the same BS may be clustered in several disconnected smaller areas. A natural question arises: would this fragmented cell association pattern result in more handovers and possibly more handover failures?

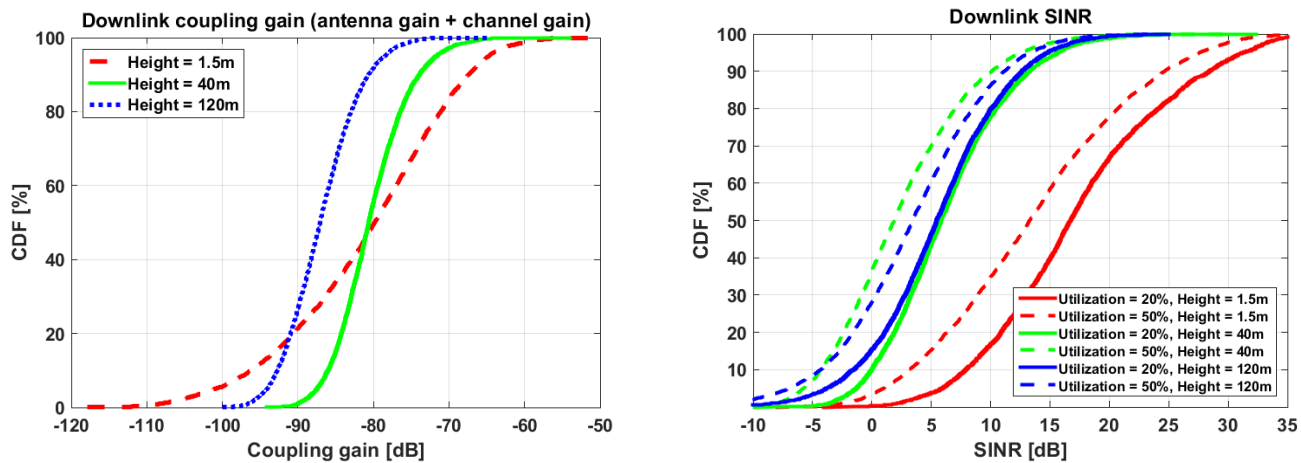

Figure 5: Downlink coupling gain and SINR distributions versus UE heights

The overall SINR level is significantly worse for aerial UEs than for UEs on the ground, as illustrated in Figure 5. The reduced SINR might lead to a higher probability of radio link failures and failed handovers. The handover command may get lost or UE may not be able to successfully connect to target cell after receiving the handover command. Also, the measurement reports which serve as input to handover decisions, might get lost or are not triggered fast enough which may delay the handover decision.

Due to the complicated factors, which depend on the scenario, BS antenna pattern as well as the trajectories and speeds of aerial UEs, it is difficult to predict the handover performance of aerial UEs. To better understand the mobility performance, the 3GPP study item on enhanced LTE support for aerial vehicles [7] has identified the study of cell selection, handover efficiency and robustness as key objectives.

*C. Aerial UE identification*

Most of the LTE connectivity optimization approaches discussed above for small UAVs are implicitly built on a basic assumption that the network can identify that (1) the UE is an airborne capable UE and (2) the airborne capable UE is flying. Here, we use the term *airborne capable UE* to generally refer to UEs that are certified or have special subscription to connect to the LTE network while airborne.

Identifying whether a UE is airborne capable or not is relatively straightforward. During the connection setup, the UE can rely on some signaling to indicate its airborne capability. The usual procedure to indicate UE capability is via radio resource control (RRC) signaling. The next question is how to identify an airborne capable UE that is flying. Perhaps the simplest solution is to request the airborne capable UE to explicitly send a mandatory message to inform the network that it is in flying mode.

Another issue is how to identify a terrestrial LTE UE that is flying. For example, this may occur when a user attaches his/her mobile device that does not have possible "LTE drone capability" to a UAV and then flies the UAV. The flying terrestrial UE may generate excessive interference to the network, and may not be allowed by regulations in some regions. Identifying the flying terrestrial UE may enable the network to take proper measures. For example, optimized performance enhancing solutions may be used if the network detects the flying terrestrial UE but decides to continue serving the UE. Alternatively, the network may limit the service or even drop the connection.

Identifying a flying terrestrial UE is a challenging task. It is an overkill to request the network to identify every terrestrial UE that may potentially be flying. From the network perspective, the main purpose of identifying flying terrestrial UEs is to prevent the excessive uplink interference that may be caused by the UEs. Therefore, it seems reasonable that the network starts to identify potential flying terrestrial UEs when the network detects the increased interference or is triggered by some other event (e.g. emergency detection).

Assuming for now that the network has detected the increased interference potentially caused by some flying UEs or triggered by some other event, it starts to identify the flying UEs. The network then can scan through the connected UEs to identify the ones that are potentially flying. The network may exploit the pattern of the received signal powers of the multiple BSs, using the fact that the uplink signal from a flying UE can reach BSs that are far away. The network may also consider estimating the position of the UE to identify if the UE is flying. Still, it may be difficult to identify whether the UE is on high floors of a high-rise building or it is flying. Speed estimation via Doppler analysis may be used, with the assumption that indoor UEs are of low mobility and flying UEs are of higher mobility. However, it is possible that the UAV is flying at a low speed and hovering over the area of operation. In summary, it is a challenging problem to identify a terrestrial UE that is flying, and deserves further study.

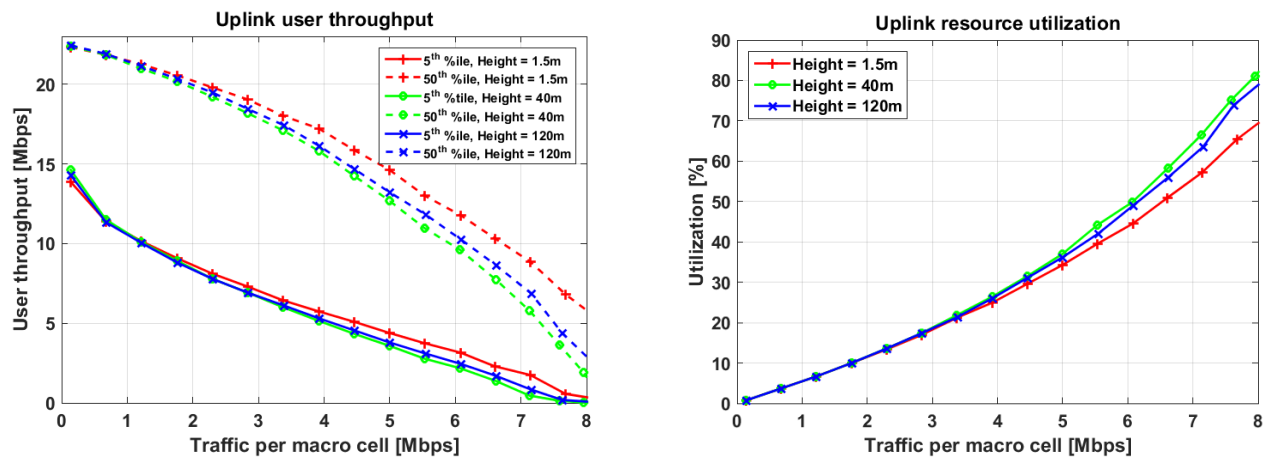

Figure 6: Uplink throughput and resource utilization versus UE heights: the ratio of the number of aerial UEs to the total number of UEs is 10%.

## VI. CONCLUSIONS AND RESEARCH DIRECTIONS

New and exciting applications for small UAVs have attracted much attention from academia, industry, and regulation bodies. Mobile networks offer wide area, high speed, and secure wireless connectivity, which can enhance control and safety of UAV operations. This article has particularly focused on LTE connectivity for UAVs, although most of the lessons herein would likely apply to other networks with UAVs. We believe that the existing mobile LTE networks targeting terrestrial usage should be able to offer wide-area wireless connectivity to the initial deployment of small UAVs. We have also identified performance enhancing solutions to optimize LTE connectivity towards more effective and efficient connectivity for small UAVs while protecting the performance of ground mobile devices. These enhancements are of importance, especially when the deployment of small UAVs gains momentum and the number of UAV connections increases.

UAV communication is an emerging and underexplored field. We conclude by pointing out some fruitful avenues for future research.

*UAV-to-UAV communication*. Collision avoidance is important for safe operation of UAVs. While coordination of UAV operations through UAV traffic management system is one way, it is also essential that UAV can detect and avoid nearby aircraft. It is of interest to explore whether LTE module onboard UAV can be reused for sensing and detection by exploiting the LTE features such as device-to-device (D2D) and vehicle-to-everything (V2X) [13]. LTE D2D/V2X features may also be used for out-of-coverage UAV communications.

*Beyond low altitude UAVs*. The focus of this article is connectivity for low altitude UAVs. A natural extension is to explore the potential of mobile network connectivity for higher altitude aircrafts, e.g., onboard mobile broadband connectivity for airlines. The link budgets need to be higher for serving high altitude aircrafts. Beamforming is an attractive technique to enable wide-area connectivity for high altitude aircrafts [14].

*5G for UAVs*. It is expected that the next generation 5G networks should have higher capacity in providing connectivity services to both terrestrial and aerial devices. The 5G study item [15] is a step towards this direction. It is our goal that new advanced technologies will be introduced in 5G networks to achieve ubiquitous mobile broadband coverage both on the ground and in the sky.

## BIOGRAPHIES

**Xingqin Lin** received his Ph.D. degree in electrical and computer engineering from the University of Texas at Austin in 2014. He is currently a researcher at Ericsson Research Silicon Valley, and leads air-to-ground communications research and standardization. He held summer internships at Qualcomm, Nokia Networks, and Bell Labs. He received the MCD fellowship from UT Austin and was recognized by Ericsson for outstanding contributions to NB-IoT. He is an Editor of the *IEEE Communications Letters*.

**Vijaya Yajnanarayana** received the MS degree in electrical engineering (EE) from the Illinois Institute of Technology, Chicago, and PhD degree in EE from the KTH Royal Institute of Technology, Stockholm, Sweden, in 2007 and 2017, respectively. He is a recipient of Program of Excellence Award from KTH Royal Institute of Technology which carried an award of 1 Million Swedish kroner. Currently, he is working as a Researcher in Ericsson Radio Research Group in Stockholm, Sweden.

**Siva D. Muruganathan** received his PhD degree in electrical engineering from the University of Calgary, Canada in 2008. He is currently with Ericsson Canada working as a Researcher and 3GPP RAN1 delegate. He previously held research/postdoctoral positions at BlackBerry Limited, CRC Canada, and the University of Alberta, Canada. His recent standardization work has been in the areas of MIMO and air-to-ground communications. He was a Rapporteur for the 3GPP Release-15 study on LTE Aerials.

**Shiwei Gao** is a researcher at Ericsson Canada, working on multi-antenna technologies and their standardization in LTE and NR since 2014. He was previously with Nortel Networks from 2007 to 2009 and Blackberry from 2009 to 2013, working on advanced wireless technology research and base station designs. His current research interests include MIMO and their application in wireless communications.

**Henrik Asplund** received the M.Sc. degree from Uppsala University, Sweden in 1996 and joined Ericsson Research, Stockholm, Sweden in the same year. Since then he has been working in the field of antennas and propagation supporting pre-development and standardization of all major wireless technologies from 2G to 5G. His current research interests include antenna techniques, radio channel measurements and modeling, and deployment options for 5G including higher frequencies.

**Helka-Liina Määttänen** received her M.Sc. and Ph.D. degree in communications engineering from Helsinki University of Technology, in 2004 and 2012, respectively. From 2006–2011, she was an external consultant at Nokia and Renesas Mobile for downlink MIMO systems. She worked at Renesas Mobile from 2011 and at Broadcom from 2013 and joined Ericsson in 2014. She has been attending 3GPP WG1 and is currently attending 3GPP WG2 as a standardization delegate.

**Mattias Bergström** is a Senior Researcher at Ericsson Research, Stockholm, Sweden. His work has been focused on radio protocols for 3GPP systems, and he has been contributing to the 3GPP standardization in the last six years. He has a M.Sc. in Wireless Communication from KTH (Royal Institute of Technology) in Stockholm, Sweden.

**Sebastian Euler** joined Ericsson Research in 2016. He has since focused on making user mobility work, both in NR as well as for unmanned aerial vehicles in LTE. He received his PhD in particle physics from RWTH Aachen University in 2014, after which held a Postdoc position at Uppsala University, and has worked with neutrino experiments in Antarctica during that time.

**Y.-P. Eric Wang** is a Principal Researcher at Ericsson Research. He holds a PhD degree in electrical engineering from the University of Michigan, Ann Arbor. Dr. Wang was an Associate Editor of the *IEEE Transactions on Vehicular Technology* from 2003 to 2007. He was a co-recipient of Ericsson's Inventors of the Year award in 2006. Dr. Wang has contributed to more than 150 U.S. patents and more than 50 IEEE articles.